\newif\ifeprint\eprinttrue
\ifeprint
\documentclass
[rmp,twocolumn,amsfonts,amsmath,showkeys,letterpaper,raggedbottom]{revtex4}
\else
\documentclass
[rmp,preprint,amsfonts,amsmath,showkeys,letterpaper,raggedbottom]{revtex4}
\fi
\date{\today}

\def\TITLE{Using binding free energy to guide ligand design}
\def\KEYWORDS{ligand design, binding free energy,
              molecular distribution, EM-algorithm, molecular symmetry}

\usepackage{times}
\usepackage[bookmarks=false]{hyperref}
\hypersetup{pdftitle={\TITLE},
pdfauthor={C. F. F. Karney; J. E. Ferrara; C. D. Spence},
pdfkeywords={\KEYWORDS}}

\ifeprint
\else
\fi
\bibpunct{[}{]}{;}{n}{}{,}

\makeatletter\def\raggedcolumn@skip{\vskip\z@\@plus.0001fil\relax}\makeatother

\def\hrefx#1#2#3{\href{#1}{#2}\penalty0\href{#1}{#3}}

\DeclareMathAlphabet{\mathbsf}{OT1}{cmss}{bx}{n}
\def\v#1{\mathbf{#1}}
\def\M#1{\mathbsf{#1}}

\def\ave#1{\langle#1\rangle}
\def\mean#1{\langle\!\langle#1\rangle\!\rangle}
\def\transp{^{\mathsf T}}

\begin{document}

\ifeprint
\noindent\mbox{\begin{minipage}[b]{\textwidth}
\begin{flushright}\tt\footnotesize
Link: \href{http://charles.karney.info/biblio/design.html}
           {http://charles.karney.info/biblio/design.html}\par\vspace{0.5ex}
E-print: \href{http://arxiv.org/abs/physics/0511061}
                             {arXiv:physics/0511061}\par
\vspace{2ex}
\end{flushright}
\end{minipage}
\hspace{-\textwidth}}
\fi

\title{\TITLE}
\ifeprint
\author{\href{http://charles.karney.info}{Charles F. F. Karney}}
\else
\author{Charles F. F. Karney}
\fi
\email{ckarney@sarnoff.com}
\author{Jason E. Ferrara}
\author{Clay D. Spence}
\affiliation{\href{http://www.sarnoff.com}{Sarnoff Corporation},
  Princeton, NJ 08543-5300}

\begin{abstract}
The molecular distributions obtained from canonical Monte Carlo
simulations can be used to find an approximate interaction energy.  This
serves as the basis of a method for estimating the binding free energy
for a ligand to a protein which enables the free energy to be used to
direct the design of ligands which bind to a protein with high affinity.

\keywords{\KEYWORDS}
\end{abstract}

\maketitle

\section{Introduction}

The goal of structure-based drug design is to create a ligand which
binds with high affinity to a protein target.  An exciting prospect is
the ability to carry out this design process computationally and thereby
obtain a series of drug leads which are potent and which can be
subsequently optimized for drug-like properties.  In an earlier paper,
we presented the wormhole method \cite{karney05a} which, subject to some
limitations, allows the binding affinity of a given drug ligand to a
protein to be computed.  For the purposes of structure-based drug
design, we might imagine using the wormhole method to screen a large
number of known molecules against the protein.  This suffers from two
serious drawbacks: only a tiny fraction of feasible drug-like chemicals
can be assessed in this way; and it is not initially known where or how
the compounds are likely to bind to the protein.

One strategy for alleviating these problems is to use a fragment-based
approach \cite{miranker91}.  Here fragments are small (usually rigid)
organic molecules.  By a judicious choice of fragments, a large and
diverse set of drug-like molecules may be built {\it in silico} by
forming bonds between them.  Because the number of fragments is small,
$O(100)$, and because they are relatively simple, we can compute maps of
where the fragments bind to the protein.  These data then serve as the
building blocks to create larger drug-like molecules.  The process of
constructing these molecules, of necessity, provides the binding mode to
the protein.  The key is, of course, to build the large molecules in
such a way as to optimize the binding affinity.

We face two challenges here.  Given a partially built candidate
molecule, can we quickly assess how a particular fragment can be added?
Having grown the molecule with the addition of a fragment, how can we
rapidly compute the resulting binding affinity to evaluate whether the
new molecule is acceptable?

In this paper we describe possible solutions to both these problems.
Let us start by giving an overall description of the method.  Our
computational model consists of a protein which is either kept rigid or
is allowed to have a small number of degrees of freedom interacting with
a ligand.  We assume that the ligand is ``made'' by connecting several
simpler organic fragments together and, for simplicity, we take the
fragments to be rigid and assume that they are joined by rotatable
bonds; however, the method is easily generalized to remove these
restrictions.  This system is described by a conventional force field
such as Amber \cite{cornell95} with the effects of the solvent captured
by an implicit solvent model such as GB/SA
\nocite{still90}\cite{still90,qiu97}.  In this model, we can limit the
number of degrees of freedom of the system to a manageable number,
$O(10)$.  For the purposes of this discussion, we assume that we have
identified a binding site on the protein.  Our goal is to design a set
of ligands (created from the fragments) which bind to the protein with
high affinity.  It would be possible to build additional criteria into
the design process, e.g., synthesizability, solubility, etc.; however,
these considerations are beyond the scope of this paper.  Our standard
for success is that the approximate techniques we develop lead to
ligands with high affinity as predicted by the full force field outlined
above.  For this purpose it is convenient to regard the full
computational model as ``exact''.  The degree of agreement with
experimental data, while crucial, entails validation of the force field
which, again, is beyond the scope of this work.

We begin by performing wormhole Monte Carlo simulations \cite{karney05a}
of each of the various fragments binding to the protein.  These
calculations give the binding affinity of each fragment to the protein
and equilibrium distributions of the fragment-protein system.  We next
fit an analytic function, a Gaussian mixture, to these molecular
distributions.  The fits for two fragments are then used for two
purposes: to find feasible ways to form a bond between the fragments,
creating a larger ligand and to give an approximate interaction energy
for the newly created ligand with the protein which, in turn, allows for
rapidly computing its binding affinity via the wormhole method.

The first part of this paper describes techniques for fitting a Gaussian
mixture to a distribution of molecular configurations.  We adopt the
well-known EM method \cite{dempster77} for this purpose; however, we
have to adapt the method to deal with two peculiarities of molecular
distributions: firstly, in the presence of constraints, the
distributions lie on a sub-manifold of Cartesian space; secondly, for
symmetric molecules, we can make a better fit by respecting the
symmetry.

Two applications of Gaussian mixtures for molecular distributions are
described next.  They may be used to define suitable portals for the
wormhole method; this provides a more robust method than the use of
ellipsoidal portals given in \cite{karney05a}.  They may also be used to
provide an approximation for the energy of a molecular system.

Finally, we describe how these tools may be combined to compute an
approximate binding affinity which allows the binding free energy to be
used to direct the design of ligands.

\section{Gaussian Mixtures}

There is often an interest in fitting some observed data with a
``model'', an analytic function which approximates the data.  One
important category of data is the set of configurations of a molecular
system given, for example, by the results of a Monte Carlo simulation.
An analytic fit then provides an approximate but compact representation
of the observed data.  Because the samples from a canonical ensemble
Monte Carlo simulation are drawn from a distribution which is
proportional to $\exp(-\beta E(\v x))$, where $E(\v x)$ is the energy of
the system in configuration $\v x$, $\beta = 1/(kT)$, $k$ is the
Boltzmann constant, and $T$ is the temperature, the analytic fit can
also be used to give an approximate expression for the energy of a
molecular configuration.

An important class of models is the mixture of Gaussians and the EM
(expectation-maximization) algorithm \cite{dempster77} is frequently
used to optimize this model based on the maximum likelihood.  We begin
by reviewing an iteration of the standard EM algorithm including the
straightforward extension of allowing the samples to have a statistical
weight.  Assume that our data is
\[
[\v x_1,\v x_2,\v x_3,\ldots,\v x_n] ,
\]
where $\v x_i$ is a point in $\mathbb R^d$ and that associated with each
of the samples $\v x_i$ is a scalar weight $w_i$.  This weight might
arise from coalescing consecutive identical samples from a Monte Carlo
simulation (because of a run of rejected moves) or because the Monte
Carlo sampling is carried out with a non-physical energy $E^*$ in which
case we have $w_i = \exp[-\beta(E(\v x_i) - E^*(\v x_i))]$.

Let the current fit be
\[
f(\v x) = \sum_{j=0}^{m-1} \alpha_j G(\v x; \v y_j, \M C_j),
\]
where $\sum_{j=0}^{m-1} \alpha_j = 1$ and $G(\v x; \v y_j, \M C_j)$ is a
$d$-dimensional Gaussian with unit volume and mean, $\v y_j$, and
covariance, $\M C_j$.  The goal is to find the set $\{\alpha_j, \v y_j,
\M C_j\}$ which maximizes the log-likelihood
\begin{equation}\label{logl}
L = \ave{ \ln f(\v x_i); w_i }_i,
\end{equation}
where $\ave{ \cdot; \cdot }_i$ denotes the weighted arithmetic mean,
\begin{equation}\label{ave}
\ave{X_i; w_i}_i = \sum_{i=1}^n w_i X_i \bigg/ \sum_{i=1}^n w_i.
\end{equation}
An EM iteration proceeds as follows:
\begin{subequations}\label{em}
\begin{eqnarray}
g_{ij} &=& \frac{\alpha_j G(\v x_i; \v y_j, \M C_j)}{f(\v x_i)},\\
\alpha_j^\mathrm{new} &=& \ave{g_{ij}; w_i}_i,\\
\v y_j^\mathrm{new} &=&
\ave{ \v x_i;  g_{ij} w_i }_i, \label{emmean}\\
\M C_j^\mathrm{new} &=&
\ave{ 
(\v x_i - \v y_j^\mathrm{new})
(\v x_i - \v y_j^\mathrm{new})\transp ;
g_{ij} w_i }_i. \label{emvar}
\end{eqnarray}
\end{subequations}
With this procedure, the log-likelihood is guaranteed to converge to a
local maximum \cite{dempster77}.  The quantity $g_{ij}$ gives the
responsibility of the Gaussian component $G(\cdot; \v y_j, \M C_j)$ for
$\v x_i$.

\subsection{Non-Cartesian space}

When simulating complex molecules it is important to reduce the
dimensionality of configuration space by imposing, for example,
constraints on bond lengths and bond angles.  For example, when
simulating biphenyl (two benzene rings linked by a single bond), the
energetics of the molecule allows us to treat it as two rigid rings
connected by a bond that permits only torsional movement.  The complete
configuration of the molecule is then given by the position and
orientation of one of the rings together with the torsion angle of the
connecting bond.

We will represent torsion angles as a point on the circle $\mathbb S^1$.
Orientations are conveniently represented as unit quaternions
\cite{karney05b}; however, because $q$ and $-q$ represent the same
orientation, orientations are defined as a pair of opposite points on
$\mathbb S^3$.  In Mardia and Jupp \cite{mardia99}, a distinction is
made between a directed line through the origin, a {\em direction}
(which can represent the torsion angle), and an undirected line through
the origin, an {\em axis} (which can represent orientations of general
molecules).  The orientation of a diatomic molecule, for example
$\mathrm{CO}$, would be given by a unit vector, i.e., a direction on
$\mathbb S^2$.  The full configuration $\v x$ is then a mixture of
Cartesian coordinates and ``angle-like'' coordinates in $\mathbb S^l$.
Our strategy for applying Gaussian fits to points in this mixed topology
is to replace eq.~(\ref{emmean}) by
\[
\v y_j^\mathrm{new} =
\mean{ \v x_i;  g_{ij} w_i }_i,
\]
where $\mean{\cdot ; \cdot}_i$ is the appropriate weighted ``physical''
mean of $\v x_i$.  For the angle-like coordinates, we find the mean by
embedding $\mathbb S^l$ in $\mathbb R^{l+1}$.  The mean direction is
given by the direction of the weighted sum of the unit vectors
\cite{mardia99}, while the mean axis is given by the axis about which
the moment of inertia of the weighted axes is minimum
\cite{mardia99,karney05b}.

Similarly, we replace eq.~(\ref{emvar}) by
\[
\M C_j^\mathrm{new} =
\ave{ 
\v d (\v x_i, \v y_j^\mathrm{new})
\v d (\v x_i, \v y_j^\mathrm{new})\transp ;
g_{ij} w_i }_i,
\]
where $\v d(\v x, \v y)$ is a displacement in $\mathbb R^d$ from $\v y$
to $\v x$.  The operation of $\v d$ is to map configurations into a
local Cartesian space centered at $\v y$.  In order to make such a
mapping for the angle-like coordinates, we project the sphere (in the
case of directions) or hemisphere (in the case of axes) onto a ball in
$\mathbb R^l$ using a generalization of the Lambert azimuthal equal-area
projection \cite{karney05a} with the pole of the projection given by the
mean.  It is important that the projection preserve area so that its
Jacobian is constant; in this way, integrals in the projected space are
the same (up to a multiplicative constant) as integrals in the original
space.

In carrying out this extension of Gaussians to non-Cartesian geometries,
we have lost an important property of the Gaussian.  In $\mathbb R^d$,
if we fit a single Gaussian to arbitrary data, then the log-likelihood
is maximized by choosing the mean and covariance of the Gaussian equal
to those of the data.  We are not aware of a generalization of the
Gaussian which preserves this property for our more complex geometries.
However, the prescription given above presumably nearly preserves this
property provided that the covariances of the individual components in
the mixture are sufficiently small that the Gaussians do not ``wrap
around'' $\mathbb S^l$ to any great degree.  We will address this issue
later.

In the discussion above, we have implicitly assumed that a uniform
measure on $\mathbb S^l$ is the natural metric for angle-like
coordinates.  This is the case for the orientation of a molecule and for
torsion angles.  However, the situation is more complex for molecules
whose bond angles can vary (for example to treat the common
conformations of cyclohexane).  A full treatment of such cases is beyond
the scope of this paper.  However, the strategy would be the same as
given here: determine a suitable mean and then map the samples to a
locally Cartesian space centered at the mean in such a way that
configuration space integrals can be expressed in the transformed space
with a constant Jacobian.

\subsection{Incorporation of symmetries}

The use of symmetries allows the simplification of many problems.  In
describing molecular configurations, we encounter both discrete and
continuous symmetries.  Examples of the latter are translational and
orientational invariance when simulating a solute molecule in a large
volume of solvent, rotational invariance about the axis in a diatomic
molecule, etc.  Such symmetries are best treated by expressing the
molecular configurations in a lower dimensional space thereby ignoring
the symmetry coordinates.  Thus the ``orientation'' of a diatomic
molecule can be expressed as a direction on $\mathbb S^2$ rather than as
an axis on $\mathbb S^3$.

Let us describe some typical discrete symmetries that arise in molecular
systems.  A molecule of methane, $\mathrm{CH_4}$, may be oriented in 12
different ways (the order of the tetrahedral group $T$) that leave like
atoms in the same positions.  We do not treat the reflection symmetry of
methane as an additional symmetry because such inversions do not occur
under normal conditions.

A more complex example is biphenyl.  When bound to a protein, this has 8
symmetries made up of combinations of $180^\circ$ rotations of the
benzene rings about the connecting bond and an interchange of the two
rings. When biphenyl is placed in any of its 8 symmetric positions the
resulting system energy and hence the equilibrium distribution is the
same.  If, on the other hand, the biphenyl is free in solution, then we
remove the continuous symmetries by fixing the position and orientation
of one of the rings.  There are then 4 symmetries given by $\psi\mapsto
\pm\psi$ and $\psi\mapsto \pm\psi + \pi$ where $\psi$ is the torsion
angle.  These correspond to rotating the free ring by $180^\circ$ and
changing the sign of the torsion angle.  The latter operation places the
biphenyl into its mirror symmetric conformation (but {\it not} by
inverting the molecule).  This symmetry is normally excluded when
biphenyl is bound to a protein, because the protein binding pocket will
not exhibit the same symmetry (because proteins are chiral).

We shall suppose that the system has a $k$-order symmetry, which can be
described by a symmetry operator $S(\cdot, l)$ where $0\le l<k$ and
\[
\v x \mapsto S(\v x, l)
\]
maps the configuration into one of the $k$ symmetric configurations.  We
can compose symmetry operations with $S( S(\v x, l'), l) = S(\v x,
l\oplus l').$ Clearly $\oplus$ defines a group of order $k$.  We will
take identity element to be $0$ and define the inverse of $l$ to be
$\bar l$ (thus, $l \oplus \bar l = 0$).

In fitting a Gaussian mixture to data, we can use $S$ both to symmetrize
the samples and to symmetrize the fit.  However, by using the properties
of $S$ the computational complexity increases only by $k$ (instead of
$k^2$).  We begin by symmetrizing the fit,
\[
f(\v x) = \sum_{j=0}^{m-1} \alpha_j
\frac1k \sum_{l=0}^{k-1} G(S(\v x, l); \v y_j, \M C_j).
\]
For simplicity, we apply the symmetry operation through the
configuration argument of $G$ rather than via $\v y_j$ or $\M C_j$.
From this definition, it is easy to show that
\[
f(S(\v x, l)) = f(\v x).
\]
(This follows from the group properties of $\oplus$.)  In forming the
responsibility matrix, we start by computing the responsibility of the
component $G(S(\cdot, l'); \v y_j, \M C_j)$ for the symmetrized data
point $S(\v x_i, l)$,
\begin{eqnarray*}
g_{iljl'} &=&
\frac1k
\frac{ \alpha_j G(S(S(\v x_i,l), l'); \v y_j, \M C_j) }
{f(S(\v x_i, l))},\\
&=&
\frac1k
\frac{ \alpha_j G(S(\v x_i,l'\oplus l); \v y_j, \M C_j) }
{f(\v x_i)},\\
&=& g_{ij(l'\oplus l)}
\end{eqnarray*}
where
\[
g_{ijl} = \frac1k
\frac{ \alpha_j G(S(\v x_i,l); \v y_j, \M C_j) }
	             { f(\v x_i)}.
\]
We can now update the components using
\begin{eqnarray*}
\alpha_j^\mathrm{new} &=& \ave{ g_{ijl}; w_i }_{i,l},\\
\v y_j^\mathrm{new} &=&
  \mean{ S(\v x_i, l);  g_{ijl} w_i }_{i,l},\\
\M C_j^\mathrm{new} &=& \ave{
\v d(S(\v x_i, l), \v y_j^\mathrm{new})
\v d(S(\v x_i, l), \v y_j^\mathrm{new})\transp;
g_{ijl} w_i }_{i,l},
\end{eqnarray*}
Here in forming $\ave{\cdot;\cdot}$ and $\mean{\cdot;\cdot}$, we sum
over $i\in(0,n]$ and $l\in[0,k)$.

\subsection{Extension of the greedy algorithm}

In the foregoing, we have supposed that the number of components in the
fit is known.  In general, this is not the case and various algorithms
have been proposed to grow the number of components in such a way that a
fit close to the global maximum for the log-likelihood is tracked.  Here
we adapt the greedy EM algorithm \cite{verbeek03} for adding components
so that symmetries can be included.  We determine the optimal number of
components by minimizing a cost function involving the minimal
description length \cite[\S7.4.2]{deco96},
\begin{equation}\label{mdl}
C = -L + 
\frac p2\biggl[m\biggl(1 + d + \frac{d(d+1)}2\biggr) - 1\biggr] \ln n .
\end{equation}
The term in brackets gives the number of free parameters in an
$m$-component fit and $p$, which is normally unity, is a parameter that
can be adjusted to penalize the addition of more components.

Let us review the greedy algorithm \cite{verbeek03}.  After the EM
algorithm has converged for an $m$-component fit, we attempt to add a
new component (with index $m$) as follows.  Initially, each data point
$\v x_i$ is assigned to the component $j$ for which $g_{ij}$ is maximum.
In this way the data is partitioned into $m$ sets $A_j$.  We make
several splits of each component $j$ by selecting two random samples
from $A_j$ and partitioning $A_j$ into two subsets based on closeness to
the two random samples.  A tentative new component is added with
$\alpha_m=\alpha_j/2$ and mean and covariance given by one of the two
subsets.  The resulting tentative fit undergoes partial EM iterations
where $\alpha_m$, $\v y_m$, and $\M C_m$ are adjusted and the $\alpha_j$
for $0\le j < m$ are merely scaled by $1-\alpha_m$ (with the
corresponding $\v y_j$ and $\M C_j$ held fixed).  This procedure is
repeated several times for each of the $m$ components and the fit with
the maximum log-likelihood (following the partial EM updates) is
selected as the $(m+1)$-component fit which is then subjected to full EM
updates.

When treating weighted samples, we modify the procedure above by
selecting the two components from $A_j$ with probabilities proportional
to their weights.  Because we do this several times, we use the Walker
algorithm \cite{walker77} to make these selections.

In order to include symmetries, we generalize $A_j$ above to $A_{jl'}$
which contains those $S(\v x_i, l)$ for which $g_{iljl'} =
g_{ij(l'\oplus l)}$ is maximum.  It is only necessary to consider
splitting the $m$ unsymmetrized components of the existing fit; thus we
only need to determine $A_{j0}$.  We can do this by assigning an
unsymmetrized sample $\v x_i$ to a symmetrized component by finding the
$j$ and $l'$ which maximizes $g_{ijl'}$, and then adding $S(\v x_i, l =
\bar{l'})$ to $A_{j0}$.

As before, we partition each $A_{j0}$ into two subsets by picking two
random samples from $A_{j0}$ (according to the sample weights) and using
the distance as defined by $\v d$ as the closeness metric.  For each
subset we use an initial $\alpha_m = \alpha_j/2$ and $(\v y_m, \M C_m)$
computed from the data in the subset.

The partial EM update then consists of
updating the responsibilities for the new component,
\[
g_{iml} = 
\frac{\alpha_m G(S(\v x_i,l); \v y_m, \M C_m)}
{(1-\alpha_m) k f(\v x_i) +
 \alpha_m \sum_{l'} G(S(\v x_i,l'); \v y_m, \M C_m)},
\]
where we need to evaluate $g_{iml}$ for $l \in [0,k)$ and for $i \in
A_{j0}$, i.e., for all $i$, for which $S(\v x_i, l)\in A_{j0}$ for some
$l$.  The update of the $m$th component is then
\begin{eqnarray*}
\alpha_m^\mathrm{new} &=& \ave{ g_{iml}; w_i }_{i^*,l},\\
\v y_m^\mathrm{new} &=&
  \mean{ S(\v x_i, l);  g_{iml} w_i }_{i^*,l},\\
\M C_m^\mathrm{new} &=& \ave{
\v d(S(\v x_i, l), \v y_m^\mathrm{new})
\v d(S(\v x_i, l), \v y_m^\mathrm{new})\transp;
g_{iml} w_i }_{i^*,l},
\end{eqnarray*}
where the subscript $i^*$ indicates that the sums over $i$ should
include only $i \in A_{j0}$.

\subsection{Loose ends}

Finding the 1-component fit with non-symmetric data is a simple matter
of computing the mean and covariance of the data.  However, if we are
performing a symmetric fit, we need to apply EM iterations to obtain a
converged one-component fit.  To determine a starting point for these EM
iterations we pick a random ``central'' sample, and transform the other
samples using the symmetry operator so that they are as close as
possible to the selected sample.  The resulting $n$ symmetry-transformed
samples are used to define a tentative $(\v y_0, \M C_0)$.  This
procedure is repeated several times with different central samples and
the $(\v y_0, \M C_0)$ which yields the maximum log-likelihood is used
as the initial guess for the first component.

The EM algorithm can fail with poorly conditioned samples.  For example,
one Gaussian component might converge to a group of samples which are in
a lower dimensional space.  We avoid this problem by placing a lower
limit on the maximum eigenvalue of the covariance matrix and by placing
a lower limit on the ratio of the minimum to maximum eigenvalues.  This
makes the EM algorithm more robust possibly at the cost of requiring
more components to maximize the log-likelihood

In Monte Carlo applications, we may wish to avoid Gaussian components
where any of the angle-like coordinates ``wrap around''.  The presence
of such wrapping may destroy detailed balance because a transition to a
wrapped sample drawn from such a Gaussian is not balanced by a reverse
process.  We can limit the effect of the wrapping by checking those
diagonal elements of the covariance matrix corresponding to the
angle-like coordinates.  If these are so large that wrapping occurs
within 3 standard deviations of the mean, for example, then we can scale
the corresponding rows and columns of the covariance matrix
appropriately so that wrapping is limited to the small fraction of
samples beyond 3 standard deviations.  Here again, the algorithm can
adjust to this constraint with additional Gaussian components.  In the
next section, we will show how detailed balance can be maintained
exactly for wormhole moves even in the face of wrapped angle-like
coordinates.

The result of a canonical Monte Carlo simulation is a set of
configurations $\v x_i$ drawn from the underlying Boltzmann distribution
proportional to $\exp(-\beta E(\v x))$, together with the corresponding
energies $E(\v x_i)$.  In the foregoing discussion, we make the fit to
the configurations, essentially ignoring the energies.  This is an
appropriate use of the data from a Monte Carlo simulation where the
sample configurations constitute the ``primary'' data.  One application
where we could make use of the energies is when making fits to several
independent Monte Carlo runs of the same system.  In this case, we can
adjust the overall weight of each independent run so that the difference
of $\-\beta \langle E(\v x_i)\rangle$ and $\langle \ln f(\v x_i)
\rangle$ is approximately the same across the runs.  (Here, $\langle
\cdot \rangle$ denotes a average over a single run.)  This adjustment is
important when the individual runs are not sufficiently long to sample
configuration space fully.

\section{Applications of Gaussian mixtures}

We use the procedure for fitting molecular distributions with a Gaussian
mixture in two ways.  The first is as a method of defining the portals
for wormhole Monte Carlo \cite{karney05a}.  In this case we are fitting
the data from several independent Monte Carlo runs and Gaussian mixtures
then offer a robust way of ``clumping'' the data with each component of
the mixture then providing a portal for the wormhole method.  The other
application provides an approximation to the energy of the system.  Here
we are more concerned about the accuracy of the fit, and we also need to
establish that the arbitrary constant that connects the energy to the
logarithm of the fit drops out when forming physically relevant
quantities.

\subsection{Portals for wormhole Monte Carlo}

The original description of wormhole Monte Carlo \cite{karney05a} was
specialized to the treatment of molecular dissociation,
\[
\mathrm A + \mathrm B \rightleftharpoons \mathrm{AB},
\]
where we sought the equilibration between the bound and unbound states
of molecules $\mathrm A$ and $\mathrm B$.  This procedure can be
generalized to deal with other types of interaction, e.g., molecular
exchange,
\[
\mathrm{AB} + \mathrm C \rightleftharpoons \mathrm{A} + \mathrm{BC},
\]
protonation,
\[
\mathrm A + \mathrm{H^+} \rightleftharpoons \mathrm{AH^+},
\]
or tautomerization,
\[
\mathrm{ABH} \rightleftharpoons \mathrm{HAB}.
\]
(In practice, the free proton in the second case would be handled by an
implicit solvent held at constant $\mathrm{pH}$.)  We therefore consider
the equilibrium of $\Lambda$ ``systems'' indexed by $\lambda$.  Each of
the systems is made of $\Phi_\lambda$ independent molecular
``complexes'' indexed by $\phi$ and each complex is made up of 1 or more
interacting molecules.  (Thus with molecular dissociation we have
$\Lambda = 2$.  The unbound system $\lambda = 0$ consists of $\Phi_0 =
2$ independent complexes, each consisting of a single molecule, $\mathrm
A$ or $\mathrm B$, while the bound system $\lambda = 1$ consists of
$\Phi_1 = 1$ complex, $\mathrm{AB}$.)

If the configuration of complex $\phi$ in system $\lambda$ is $\v
x_{\lambda\phi}$, then the full phase space is given by $\Upsilon =
\{\lambda; \v x_{\lambda0}, \v x_{\lambda1},\ldots, \v
x_{\lambda\Phi_\lambda}\}$.  Here, we have added a set of ``ignorable''
coordinates, $\v x_{\lambda0}$; the energy of the system, and hence the
equilibrium distribution function, is strictly independent of these
coordinates.  For example, in simulating a molecule in a solvent bath,
$\v x_{\lambda0}$ would include the position and orientation of the
molecule; or, when a molecule is deprotonated, it would include the
coordinates of the ``missing'' proton.  Inclusion of these ignorable
coordinates is dictated by the requirement that $\Upsilon$ span the same
phase space volume for each $\lambda$.  In practice, we do not keep
track of $\v x_{\lambda0}$, because the integral over this coordinate is
trivial (the integrand is constant!)  and we write
\[
\int d\v x_{\lambda0} = v_{\lambda0}.
\]

Wormhole Monte Carlo moves allow the state to switch between different
systems preserving detailed balance.  This allows the determination of
the ratios,
\[
W_0
\mathbin:
W_1
\mathbin:
W_2
\mathbin:
\ldots,
\]
where $W_\mu$ is the statistical weight of system $\mu$,
\[
W_\mu = \exp(-\beta F_\mu) =
\int \delta_{\lambda\mu}\exp(-\beta \Upsilon)\,d\Upsilon
,
\]
and $F_\mu$ is its free energy.  In particular, in the case of
protein-ligand binding, the dissociation constant is given by
\[
K_d = \frac 1{V_0} \frac{W_0}{W_1},
\]
where $V_0$ is the system volume.

In this more general framework, the wormhole move \cite{karney05a} is
defined as follows.  We define a set of ``portal functions,'' $w$, $w'$,
$w''$, \ldots, on $\Upsilon$, with properties
\begin{eqnarray*}
&0\le w(\Upsilon) \le 1/v < \infty,\\
&\int d\Upsilon\, w(\Upsilon) = 1,
\end{eqnarray*}
where $v$ is a representative phase-space volume of the portal function.
A wormhole move consists of the following steps: select a pair of
portals $(w, w')$ with probability $p_{ww'}$; reject the move with
probability $1 - v w(\Upsilon)$, where $\Upsilon$ is the current state;
otherwise, with probability $v w(\Upsilon)$, pick a configuration
$\Upsilon'$ with probability $w'(\Upsilon')$; and accept the move to
$\Upsilon'$ with probability
\begin{equation}\label{pacc}
P_{ww'}(\Upsilon,\Upsilon') =
\min\biggl(1, \frac{p_{w'w}}{p_{ww'}}
\frac{\exp(-\beta E^*(\Upsilon'))}
{\exp(-\beta E^*(\Upsilon))}
\frac{v'}{v}\biggr),
\end{equation}
where $E^*(\Upsilon) \approx E(\Upsilon)$ is the ``sampling'' energy of
configuration $\Upsilon$, which is used also for the conventional Monte
Carlo moves (within a system).  We term $w$ and $w'$ the source and
destination portals, respectively.  The test involving $w(\Upsilon)$
determines whether the current configuration is ``in'' the source
portal---note, however, that this test is ``fuzzy''.  If the test
succeeds, a move is attempted to the destination portal, and the move is
accepted according to a standard Boltzmann factor modified by the ratio
of the portal volumes.  In the limit of a long Markov chain, we then
have
\[
W_\mu \rightarrow C \ave{ \delta_{\lambda\mu}
\exp(-\beta [ E(\Upsilon) - E^*(\Upsilon)])},
\]
where $C$ is independent of $\mu$ and $\ave\cdot$ is the average over
the Markov chain.

Although the choice of portal functions is arbitrary, the method is only
effective if $vw(\Upsilon)$ is sufficiently large to allow wormhole
moves.  For simplicity, we restrict each portal function to a particular
system $\lambda = \mu$.  Because the complexes making up a system are
independent it is natural to consider $w(\Upsilon)$ as product of
density functions for each complex.  Thus the typical portal function is
\[
w(\Upsilon) = \delta_{\lambda\mu} \frac1{v_{\mu0}}
\prod_{\phi=1}^{\Phi_\mu} w_{\mu\phi}(\v x_{\mu\phi}),
\]
where
\[
\int w_{\mu\phi}(\v x_{\mu\phi})\,d\v x_{\mu\phi} = 1,
\]
and the factor $1/v_{\mu0}$ arises from an implicit constant density in
$\v x_{\mu0}$.  For a particular complex $\phi$ in system $\lambda$ we
need to determine a set of portal functions $w_{\mu\phi}(\v
x_{\mu\phi})$, $w'_{\mu\phi}(\v x_{\mu\phi})$, $w''_{\mu\phi}(\v
x_{\mu\phi})$, \ldots, which reflect the probable configurations for
this complex.  We obtain these portal functions using the results of
several conventional canonical Monte Carlo runs on the complex.  We make
a Gaussian fit to the resulting sets of configurations.  If the fit
contains $m$ components, then we obtain $m$ portal functions, indexed by
$j$, for this complex each of which is a symmetrized Gaussian of the
form
\begin{equation}\label{symportal}
w_{\lambda\phi j}(\v x_{\lambda\phi}) = 
\frac1{k_{\lambda\phi}} \sum_{l=0}^{k_{\lambda\phi}-1}
G(S_{\lambda\phi}(\v x_{\lambda\phi}, l);
\v y_{\lambda\phi j}, \M C_{\lambda\phi j}),
\end{equation}
where $k_{\lambda\phi}$ is the symmetry order for the complex,
$S_{\lambda\phi}$ is the corresponding symmetry operator, etc.  We take
the ``volume'' of this portal function to be
\[
v_{\lambda\phi j} =
k_{\lambda\phi}/
G(\v y_{\lambda\phi j}; \v y_{\lambda\phi j}, \M C_{\lambda\phi j}).
\]

We assume that the choice of source and destination portals is
independent so that the portal probability $p_{ww'}$ can be factored
into probabilities for $w$ and $w'$; furthermore, we assume that these
probabilities may in turn be factored into choices for the source and
destination systems and for the portals for the respective complexes for
each system.  In this case, the wormhole moves can implemented as
follows. Pick a source portal system $\mu$; if $\lambda\ne\mu$, the move
fails; otherwise consider each complex in the system $\mu$ in turn; for
complex $\phi$, pick a random portal function $j$ and pick a random
symmetry index $l$; with probability
\[
1 - v_{\mu\phi j}G(S_{\mu\phi}(\v x_{\mu\phi}, l);
\v y_{\mu\phi j}, \M C_{\mu\phi j})/k_{\mu\phi},
\]
reject the move; if none of these tests cause the move to be rejected,
the ``in'' test succeeds and we proceed with choosing the destination
portal by picking the destination system $\mu'$, picking a portal
function $j'$ and a symmetry $l'$ for each complex $\phi'$, and setting
the configuration for the complex to $S_{\mu'\phi'}(\v x_{\mu'\phi'},
l')$ where $\v x_{\mu'\phi'}$ is selected from $G(\v x_{\mu'\phi'}; \v
y_{\mu'\phi' j'}, \M C_{\mu'\phi' j'})$.  In evaluating the acceptance
probability, eq.~(\ref{pacc}), we express $p_{ww'}$ as the product of
the individual probabilities (of selecting source and destination
systems and of selecting particular portals for the source and
destination complexes).  Similarly the volume of the portal is given by
the product of $v_{\mu 0}$ and the volumes of the portal functions for
the separate complexes.

In this formulation, the symmetry of a complex is incorporated into the
portal function, eq.~(\ref{symportal}).  However the test for being in
the portal and the operation of selecting a configuration from it are
decomposed into picking a random symmetry (with equal probabilities)
followed by a test or selection on a unsymmetrized Gaussian.

If, when sampling from the destination portal, any of the angle-like
coordinates are wrapped around, then we immediately reject the whole
move.  This is necessary in order to maintain detailed balance, because
the test on the source portal never involves wrapped coordinates.  This
effectively replaces the Gaussians in the definition of the portal
functions by clipped versions which evaluate to zero for wrapped
coordinates.

There is a great deal of flexibility in the choice of portal
probabilities offered by the scheme outlined here.  Because the test of
being in the source portal is typically very inexpensive, it is
desirable to arrange that the source portal probabilities are roughly
equal.  In addition, we usually adjust the ratio of conventional to
wormhole moves so that, on average, each configuration is tested against
all the portals for every attempted conventional move.  On the other
hand, the probabilities for the destination portals would usually be
adjusted to reflect the statistical weight of the portal.

Let us turn to the details of making the Gaussian fits to define the
wormhole portals.  Because the individual Monte Carlo runs performed for
each complex are independent, it is natural to consider scaling the
overall weight to the results from each run in order to match the energy
samples.  In practice, this procedure results in rather poor fits with
too many components.  In this application, Gaussian fitting may viewed
merely as a robust clumping technique and for this purpose if suffices
to attach the same weight to all the samples.  For the same reason, we
increase $p$ to 5 in eq.~(\ref{mdl}) so that a smaller number of
components is used to make the fit.

Gaussian portals offer advantages over the use of ellipsoidal portals
proposed in \cite{karney05a}.  With a given number of components, the EM
method does a ``global'' optimization and is thus likely to obtain a
better fit and than the somewhat {\em ad hoc} scheme for choosing
ellipsoids.  Also the Gaussian fit to a configuration of independent
complexes naturally factors into a product of Gaussians for each
complex.  Thus Gaussian portals reflect the independence of complexes
properly.  Gaussian portals, combined with the mean energy for a portal
(which can be readily estimated from the energies of the samples) also
offer a rough {\em a priori} estimate of $W_\lambda$.  This, in turn,
allows us to adjust $v_{\lambda0}$ to maximize the probability of
transitions between systems and hence to reduce the error in the
eventual estimate of $W_\lambda$.  In the case of ligand-protein
binding, where we simulate single ligand and protein molecules in a
system of physical volume $V_0$, this procedure entails adjusting $V_0$
so that the fraction of time the molecules are associated is roughly
$\frac12$.

Finally, we remark that when performing a conventional Monte Carlo move
for a particular system, it is preferable to select randomly a single
complex to move.  This will result in a higher acceptance rate compared
to attempting to move all the complexes simultaneously.

\subsection{Obtaining the energy from the fit}

The result of a wormhole Monte Carlo simulation is a set of
configurations sampled from $\exp(-\beta E^*(\Upsilon))$.  If we fit an
analytic function to these samples weighted by $\exp(-\beta [E(\Upsilon)
- E^*(\Upsilon)])$, then the fit can serve as a basis for approximating
$E(\Upsilon)$.  Here we detail how we can use Gaussian mixtures to carry
out this fit and we show how to obtain approximations for the energies
of the individual complexes and how the arbitrary offsets for energies
cancel whenever energy differences are computed using the approximate
energies.

We begin by making normalized fits, $f_{\lambda\phi}$, to all the
complexes in all the systems. If the same complex appears in multiple
systems, the samples may be aggregated in order to permit a fit using
more data.  The energy of the system is taken to be the sums of the
energies of the complexes, i.e.,
\[
E(\Upsilon) = \sum_{\phi=1}^{\Phi_\lambda} E(\v x_{\lambda\phi}),
\]
and similarly for $E^*(\Upsilon)$.  The sampled configurations for each
complex are assigned weights of $\exp(-\beta [E(\v x_{\lambda\phi}) -
E^*(\v x_{\lambda\phi})])$.

The normalized fit for a particular system is then given as the product
of the fits for the contributing complexes, multiplied by $1 /
v_{\lambda0}$ and these can be combined weighted by $W_\lambda$ to
provide a fit in $\Upsilon$ space as
\[
f(\Upsilon) =
\frac{W_\lambda}W\frac1{v_{\lambda0}}
\prod_{\phi=1}^{\Phi_\lambda} f_{\lambda\phi}(\v x_{\lambda\phi}),
\]
where $W=\sum_\lambda W_\lambda$, and we have
\[
-\beta E(\Upsilon) \approx D + \ln f(\Upsilon),
\]
where $D$ is an arbitrary adjustable constant.  This provides an
approximation of the energy of a system.  In addition, we can
approximate the energies of the individual complexes by
\[
-\beta E(\v x_{\lambda\phi}) \approx D_{\lambda\phi} +
\ln f_{\lambda\phi}(\v x_{\lambda\phi}),
\]
where $D_{\lambda\phi}$ are adjustable constants which satisfy
\[
\sum_{\phi=1}^{\Phi_\lambda} D_{\lambda\phi} = D + \ln(W_\lambda/W) -
\ln v_{\lambda0},
\]
for all $\lambda$.

Let us apply this to the case of molecular dissociation.  The $\lambda =
0$ (resp.~$\lambda = 1$) system contains two complexes (resp.~one
complex) each of which is free to move within a system of 3-dimensional
volume $V_0$; thus, we have $v_{00} = (\sigma V_0)^2$ (resp.~$v_{10} =
\sigma V_0$), where $\sigma$ is the volume of orientation space.  In
this case, there is one constraint on the choice of energy offsets for
the fit energies, namely
\[
D_{10} = D_{00}+D_{01} - \ln(K_d/\sigma).
\]
As expected, this constraint does not involve $V_0$.  It is also
apparent from the form of this constraint, that differences in energy
between the unbound and bound systems will be independent of the choice
of offsets.

\section{Molecular design}

We now have the tools to tackle ligand design.  We start the process by
computing the binding affinity of the fragments.  Fragment-based design
works on the principle of building a complex molecule from simpler
sub-components.  We extend this idea by also computing the binding
affinity of the larger molecule using data from the calculation of the
binding affinity of the simpler molecules.  Finally, we describe the
process by which simple molecules can be combined.

\subsection{The single fragment binding affinity}

Our starting point is a protein target for which we know the structure
and a set of simple organic fragments.  The symmetries of the fragment
are determined.  In the case of a rigid fragment, this consists of the
set of 3-dimensional rotations which leave the molecule invariant.  The
energy of the system is computed using a conventional force field with
an implicit solvent model as described in the introduction.  The
simulation is focused on a certain portion of the protein by adding a
restraint energy which is zero if the fragment is within a region of
interest on the protein (e.g., within a binding site) and increases
parabolically outside this region.  It is possible to define the
restraint region to include a few solvent layers about the entire
surface of the protein---but this obviously results in a longer
simulation.  Including a parabolic portion to the restraint potential
allows the fragment distribution to fall off gradually and this allows
the distribution to be fit with fewer Gaussian components than with a
hard restraint.  The binding affinity of a molecule will be only weakly
dependent on the precise extent of the restraint region providing that
it encompasses the true binding site of the protein.

We perform a wormhole calculation to find the binding affinity and to
provide the distribution of fragments.  In order to determine initial
portals for this calculation we systematically search for plausible
binding modes by inserting the fragment randomly into the restraint
region with a random orientation and random conformations for the
protein and fragment (if these molecules are flexible).  This process is
most efficiently carried out with a tailored restraint region (which
prevents attempts to insert the ligand within the protein) followed by a
quick steric check (where configurations are rejected as soon as two
clashing atoms are found).  This can be followed by a crude energy
minimization using the vacuum energy model.  We can make an estimate of
how many probes need to be made in order to explore the surface of the
protein thoroughly and so to find all possible binding pockets.  This
estimate is based on the volume of the restraint region and the typical
length and orientation scales for energy variation.  A similar exercise
is carried out for the unbound system---this merely consists in finding
allowable conformations of the protein and ligand.  We then drop any of
the bound configurations whose energy exceeds the minimum unbound
energy.  The resulting bound and unbound configurations are used as
starting points for a set of conventional Monte Carlo runs with the full
sampling energy.  The initial portion of each run should be discarded
and any bound run whose energy is stuck close to (or above) the unbound
energy should be eliminated.  The resulting data from these Monte Carlo
runs is then used to determine wormhole portals using a Gaussian mixture
and to estimate a starting value of the system volume $V_0$.

In addition, we can add ``catch-all'' portals for the unbound and bound
systems.  For the unbound system, this will allow the molecules to
assume arbitrary conformations (subject to whatever constraints are
imposed by the molecular model).  For the bound system, the molecules
would be allowed to assume arbitrary conformations and in addition the
ligands would be selected from the restraint region with an arbitrary
orientation.  These catch-all portals allow new binding modes to be
discovered.

The binding affinity is then calculated using the wormhole Monte Carlo
method.  During the course of this simulation, $V_0$ is adjusted to
maintain $W_0 \sim W_1$ and if $V_0$ is increased (resp. decreased) we
reduce $W_1$ (resp.~$W_0$) by the same factor.  We may also find that
the ligand becomes trapped in a local energy well.  Whenever this is
detected (by the absence of successful wormhole moves), new Gaussian
portals are found by rerunning the Gaussian fit adding recent
configurations and restarting the binding affinity calculation.

The process converges with a sufficiently long run without the need to
add new portals and with sufficiently frequent wormhole moves between
the bound and unbound systems.  This process provides an estimate of the
dissociation constant $K_d$ for this fragment-protein interaction and a
set of samples for the bound and unbound configurations.  From this
configurational data (weighted to reflect the difference between the
full and sampling energies), we fit a Gaussian mixture to the bound and
unbound distributions.  This provides an approximation to the energy of
the protein and the ligand either unbound or as a bound complex.

\subsection{Approximate energy of combined molecules}\label{approxen}
 
Let us consider the case where we have identified two possible ligands
$\mathrm{AB}$ and $\mathrm{BC}$ and we wish to combine these via the
``overlap'' portion $\mathrm B$ to form a ligand $\mathrm{ABC}$.  We
might form a $N$-fragment ligand with $\mathrm A$ and $\mathrm C$ being
fragments and $\mathrm B$ being an $(N-2)$-fragment overlap ligand.
Alternatively, we might take $\mathrm B$ to be null and merely add a
fragment $\mathrm C$ to a $(N-1)$-fragment ligand $\mathrm A$.  (In
either case, we form a 2-fragment ligand by taking $\mathrm A$ and
$\mathrm C$ to be fragments and $\mathrm B$ to be null.)

Because we are concerned here with the energies of different
combinations of fragments, we adopt a notation for the energy where we
do not explicitly specify the molecular configuration and where
$E(\mathrm Z)$ is the energy for a single molecule $\mathrm Z$ and
$E(\mathrm X, \mathrm Y)$ is the energy for the two interacting
molecules $\mathrm X$ and $\mathrm Y$.  We write
\begin{subequations}\label{enadd}
\begin{eqnarray}
E(\mathrm{ABC})
	      &=& E(\mathrm{AB}) + E(\mathrm{BC})
                - E(\mathrm{B})
\nonumber\\ &&\qquad{}
+ \delta E(\mathrm{ABC}),
\label{enadd0}\\
    E(\mathrm{ABC}, \mathrm{P}) &=&
  E(\mathrm{AB}, \mathrm{P}) + E(\mathrm{BC}, \mathrm{P})
 - E(\mathrm{B}, \mathrm{P})
\nonumber\\ &&\qquad{}
 + \delta E(\mathrm{ABC}, \mathrm{P}),\label{enadd1}
\end{eqnarray}
\end{subequations}
where we expect $\delta E$ to be small if the energies are approximately
additive.  In this and subsequent equations, we understand the
configuration of the molecules to be consistent throughout the equation,
e.g., $\mathrm A$ is in the same configuration in all the terms.  If
$\mathrm B$ is null, then we can write $E(\mathrm{B}, \mathrm{P}) =
E(\mathrm{P})$.

We now make two approximations.  We assume that the energies involving
the simpler molecules $\mathrm{AB}$, $\mathrm{BC}$, and $\mathrm{B}$,
are given by fits to configurations from prior binding affinity
calculations and we assume that $\delta E(\mathrm{ABC}, \mathrm{P})$ and
$\delta E(\mathrm{ABC})$ are small.  If these assumptions hold, then
eq.~(\ref{enadd}) provides a method of computing the energy of the more
complex molecule $\mathrm{ABC}$ very rapidly which, in turn, allows its
binding affinity to the protein to be determined quickly.  We expect the
neglect of $\delta E(\mathrm{ABC}, \mathrm{P})$ and $\delta
E(\mathrm{ABC})$ to be most easily justified when the overlap portion
$\mathrm B$ is as large as possible; i.e., when two $(N-1)$-fragment
ligands are combined to form an $N$-fragment ligand.  In carrying out
this calculation, the arbitrary constants that enter when converting the
fits to energies cancel when considering energy differences between the
bound and unbound systems.

In our initial implementation, we build ligands by adding fragments one
at a time with no overlap portion (i.e., $\mathrm B$ is null).
Furthermore the approximate expressions for the energy are applied
recursively so that the energy of a $N$-fragment ligand interacting with
the protein is found by summing each of the individual fragment-protein
energies.

The accuracy can be improved as follows: Assume that we have computed
the binding affinity of the best ligands with up to $N-1$ fragments
using the full energy.  Fits to the distributions of these molecules
provide the corresponding approximate energies which can be used to
compute the approximate energy, using eq.~(\ref{enadd}), for
$N$-fragment ligands either by adding a single fragment or, preferably,
by using an $(N-2)$-fragment overlap.  This allows us to compute
approximate values for the binding affinity of the $N$-fragment ligands.
The binding affinity of those ligands with the best approximate binding
affinities can then be recomputed using the full energy.  Because this
latter binding affinity calculation is carried out following the similar
calculation with the approximate energy, we can use Gaussian fits to the
samples from the approximate calculation to define the initial portals
for the wormhole method with the full energy.  This procedure can be
repeated to create ligands with an arbitrary number of fragments.

In order to compute the full energy of the enlarged molecule, we may
have to determine the Amber atom types afresh, for example, using the
GAFF rules \nocite{wang04}\cite{wang04,wang05}.  In addition, we need to
determine the partial charges for the new ligand.  One simple
prescription is as follows: when two fragments, $\mathrm A$ and $\mathrm
C$ are combined to form a 2-fragment molecule $\mathrm{AC}$, the charge
on the hydrogen removed from $\mathrm A$ is donated to the derivatized
atom on $\mathrm C$ and {\em vice versa}.  This rule, which maintains
charge neutrality, can be readily generalized for larger ligands.  More
realistic charge models could be employed, if necessary.  For example,
VC/2003 \cite{gilson03} allows partial charges to be computed without
the need for a quantum calculation, while AM1-BCC
\nocite{jakalian00}\cite{jakalian00,jakalian02} gives the charge on the
basis of an relatively inexpensive AM1 quantum calculation.  The overall
expense of these more detailed charge calculations could be reduced by
carrying them out only for the ligands with the best predicted binding
affinity.  Techniques for making constrained moves of a molecule made up
of rigid fragments and for efficiently evaluating the resulting energy
(including the solvation free energy) are given in \cite{karney05c}.

We can further improve the accuracy of the approximate energy evaluation
by including some contributions to $\delta E$.  For example, we can set
$\delta E(\mathrm{ABC}, \mathrm{P}) \approx \delta E(\mathrm{ABC})$ and
evaluate $\delta E(\mathrm{ABC})$ using eq.~(\ref{enadd0}) together with
a direct evaluation of $E(\mathrm{ABC})$ (which, typically, is fast
because it does not involve the interaction with the large protein).  In
this way, we expect to include the main intra-molecular contributions to
the energy including the effects of atom removal and charge
redistribution.  In this approximation, we still neglect three-body
effects which enter into the solvation energy for the bound system,
e.g., the modification of solvation energy of $\mathrm A$ interacting
with $\mathrm P$ due to the presence of $\mathrm C$ \cite{still90}.
Also neglected are the effects of atom removal and charge redistribution
on the ligand-protein energy.

We might make a further simplification to $\delta E(\mathrm{ABC})$ by
including only some terms in energies in eq.~(\ref{enadd0}).  For
example, we might include just the torsion energy of the inter-fragment
bonds and a ``steric'' energy, which is infinite if non-bonded atoms in
$\mathrm{ABC}$ overlap and is zero otherwise.

Finally, note that we do not need to include the chemical bond energies
when forming $\mathrm{ABC}$ because these energies are the same in the
bound and unbound systems, and so cancel in the computation of the
binding affinity.

\subsection{Combining molecules}\label{combine}

When forming a complex ligand $\mathrm{ABC}$ from ligands $\mathrm{AB}$
and $\mathrm{BC}$, we need to generate starting configurations for
$\mathrm{ABC}$ for the purposes of identifying the wormhole portals
\cite{karney05a}.  We consider the problem for the bound ($\lambda = 1$)
case; the unbound case ($\lambda = 0$) follows as a straightforward
simplification.

We compute Gaussian mixtures for the configurations of $\mathrm{AB}$ and
$\mathrm{BC}$ bound, respectively, to the protein.  We draw several
sample configurations $[\Gamma_\mathrm A, \Gamma_\mathrm B,
\Gamma_\mathrm P]$ and $[\Gamma_\mathrm B', \Gamma_\mathrm C,
\Gamma_\mathrm P']$ from each possible pair of Gaussian components
selected from the two fits.  Here $\Gamma_\mathrm M$ denotes the
configuration of molecule $\mathrm M$.  We form a ``bond'' constraint
term,
\[
D^2 =
\left\|\Gamma_\mathrm B - \Gamma_\mathrm B'\right\|^2 +
\left\|\Gamma_\mathrm P - \Gamma_\mathrm P'\right\|^2 ,
\]
where $\left\|\Gamma_\mathrm M - \Gamma_\mathrm M'\right\|$ is some
suitable measure of the separation of the two configurations of molecule
$\mathrm M$ and we take $D\ge 0$.  If $\mathrm B$ is null, then
$\left\|\Gamma_\mathrm B - \Gamma_\mathrm B'\right\|^2$ is replaced by a
constraint term for the new bond between $\mathrm A$ and $\mathrm C$,
for example, an appropriately weighted sum of the squared deviations of
the bond length and bond angles from their ideal values.  In forming
$D$, we weight the various contributions so that $D$ is an approximate
distance that atoms must move to satisfy $D=0$

We now seek nearby configurations for the two systems $[\Gamma_\mathrm
A, \Gamma_\mathrm B, \Gamma_\mathrm P]$ and $[\Gamma_\mathrm B',
\Gamma_\mathrm C, \Gamma_\mathrm P']$ for which the bond constraint is
zero.  This is accomplished by gradually decreasing a ``target''
constraint, $D_t$, from the initial value of $D$ to zero.  For a
specific $D_t$, we randomly perturb the configurations in such a way as
to meet the target constraint, $D\le D_t$, and accept the new
configuration with a Boltzmann probability $\min[1, \exp(- \Delta
E/(kT'))]$, where $\Delta E$ is the change in the (fit) energy and $T'$
is an annealing temperature.  In this way, we attempt to minimize the
energy of the combined system subject to the bond constraint.  This
procedure is attempted at the physiological temperature $T' = T$ and
then at successively higher temperatures until either we achieve $D = 0$
or an upper temperature, $T' = 2T$, is reached.

If the bond constraint can be satisfied and if we include the steric
term for $\mathrm{ABC}$ in the approximate binding affinity calculation,
we repeat the above procedure to satisfy a steric constraint $S = 0$,
where $S$ measures the degree of overlap between the non-bonded atoms of
$\mathrm{ABC}$.  In this case, we perturb the molecule $\mathrm{ABC}$
subject to the constraint $D=0$ in order to reduce $S$ to zero following
a similar strategy as that used to meet the bond constraint.

If the bond and (if applicable) steric constraints can be satisfied in
such a way that both $[\Gamma_\mathrm A, \Gamma_\mathrm B,
\Gamma_\mathrm P]$ and $[\Gamma_\mathrm B, \Gamma_\mathrm C,
\Gamma_\mathrm P]$ are within a few standard deviations of one of the
components of their respective Gaussian mixtures, then $[\Gamma_\mathrm
A, \Gamma_\mathrm B, \Gamma_\mathrm C, \Gamma_\mathrm P]$ gives a
configuration for $\mathrm{ABC}$ bound to $\mathrm P$ which serves as
one of the starting points for finding portals for the bound system.

The scheme described above is appropriate when we have direct
calculations of the fit energies of $\mathrm{AB}$ and $\mathrm{BC}$
interacting with the protein.  However, if these are given by summing
the contributions over fragments, then the bond minimization of the new
bonds is carried out allowing all the inter-fragment bonds to relax.  In
this case, the ``old'' inter-fragment bonds would start in the ideal
constrained state; however, in allowing the new bonds to relax, the old
bonds are allowed to stretch so that the ligand can find a good pose
where all the inter-fragment bonds meet the constraints.

\subsection{Implementation}

The preceding sections describe the physical basis for using binding
free energy to direct ligand design.  The implementation entails
additional challenges.  There is a need for bookkeeping to associate a
molecule with the smaller molecules out of which it was created.  Care
must be taken to match up the configurations of the molecules in order
to implement the approximate energy evaluations.  In order to avoid
building the same molecule multiple times (e.g., putting the fragments
together in a different order), we use the USMILES representation
\cite{weininger89} as a unique tag for the molecule.  (Unfortunately,
this representation has shortcomings for our purposes.  It does not, in
fact, provide a unique representation of a molecule; e.g.,
$\mathrm{C1C2CC2CC3CC13}$ and $\mathrm{C1C2CC3CC3CC12}$ are two
different USMILES representations of the same molecule.  Furthermore,
USMILES only deals with the 2-dimensional structure of a molecule and
for the purposes of binding affinity, stereoisomers should be treated as
distinct.)

In our current implementation, we build molecules by adding one fragment
at a time.  The approximate binding affinity of the molecules is
computed by summing the fit energy of the individual fragments and
including the steric energy.  Any intermediate molecule meeting a
threshold binding affinity is recorded and is used as a base from which
to build larger molecules (up to a given size) in a depth-first fashion.

We have tested this procedure by building ligands which bind to
botulinum neurotoxin type B \cite{hanson00} starting with 35 organic
fragments.  A subsequent evaluation of the binding affinity using the
full energy shows that good agreement with the approximate binding
affinity in the case of 2-fragment ligands, with 90\% of the pairs
having an approximate binding affinity within 1--2 log units of the full
binding affinity.  However the agreement is poor for ligands made up of
3--5 fragments.  We attribute this to basing the approximate energy of
$N$-fragment ligands purely on the energies of single fragments.  The
errors in the use of the approximate energies may become excessive so
that the approximate binding affinity is no longer close to the full
binding affinity.  Alternatively, it's possible that the approximate
energy is still reasonably accurate but that the binding mode is
slightly wrong so that using the approximate distributions to provide
the initial portals for the full binding affinity calculation may be
inadequate; if the binding mode is quite tight, the full binding
affinity calculation may never find it.  Both of these problems would be
largely overcome by basing the approximate energy for $N$-fragment
ligands on the fit energy for $(N-1)$-fragment ligands.

\section{Discussion}

We have described a way to determine the approximate binding affinity of
a ligands based on knowledge of the binding affinity of simpler ligands
and the associated equilibrium distributions.  This procedure correctly
accounts for the loss of entropy associated with connecting molecules
together to form a larger molecule and allows the binding free energy to
be used to direct the design of ligands.

The predictive capability of our initial implementation (described
above) is limited to ligands made up of just 2 fragments.  However, we
believe that this limitation could be removed by computing the full
binding affinity of the best $(N-1)$-fragment ligands and using this as
the basis of building $N$-fragment ligands.  Such a scheme would allow a
breadth-first search which would allow the search to be directed toward
the molecules with the greatest binding affinity.

In pruning intermediate molecules, we may wish to retain
$(N+1)$-fragment molecules which have a worse binding affinity than
their $N$-fragment parents, with the expectation that this would enable
us to build better $(N+2)$-fragment molecules.  This would allow
non-functional linkers (with no intrinsic propensity to bind to the
protein) to bridge between high-affinity functional groups.

The technique described here covers joining fragments in a simply
connected fashion where each added fragment attaches at one point.  It
would be possible to generalize the method to allow the formation of
rings.  For example we might overlap the molecules $\mathrm{ABC}$,
$\mathrm{BCD}$, $\mathrm{CDA}$, $\mathrm{DAB}$ to create a 4-fragment
ring $\mathrm{(ABCD)}$.

The wormhole technique and the ability to fit distributions of molecular
configurations with Gaussian mixtures intrinsically depends on the
system having a ``small'' number of degrees of freedom, because we
require that phase space be spanned by a reasonable number of samples.
This limits the degree of flexibility that can be allowed for the
protein and dictates the use of an implicit solvent model.  On the other
hand, because we are just interested in describing where samples are
concentrated, one might expect the method to continue to function well
as the number of degrees of freedom is increased to, say, 20.

Our ability to obtain realistic results is also limited by the accuracy
of the force field and the solvent model.  There are several areas of
concern.  The force field, the methods for determining partial atomic
charges, and the solvent models have all been developed largely
independently, and it's not clear how consistent these models are.  It
is also noteworthy that the data validating the GB models is based on
comparisons with solutions to the Poisson-Boltzmann equation
\cite{feig04}, which requires specification of the atomic radii, or is
based on comparisons with experimental data for the absolute solvation
energy (from vacuum to solution) \cite{qiu97}.  More relevant for our
purposes would be a comparison against experimental data for the changes
in solvent free energy on molecular association.  Another area of
uncertainty is the charge state of the protein, which may have a small
effect when differences in binding free energy are being computed (e.g.,
with the free energy perturbation method) but which may have a large
effect on the absolute binding free energy.  The impact of salt effects
is easily incorporated into GB models \cite{srinivasan99}.  More
interesting would be a principled treatment of the protonation of
charged residues in the protein, proton exchange between the protein and
ligand, and tautomerization of the protein or the ligand.  The wormhole
method offers a natural vehicle for such a treatment avoiding the need
to treat protonation as a continuous process \cite{lee04} and avoiding
the need to add an uncharged ghost proton \cite{mongan04}.

\section*{Acknowledgment}

This work was supported by the
\hrefx{https://mrmc-www.army.mil}{U.S. Army Medical Re-}{search and Materiel
Command} under Contract No.\ DAMD17-03-C-0082.  The views, opinions, and
findings contained in this report are those of the author and should not
be construed as an official Department of the Army position, policy, or
decision.  No animal testing was conducted and no recombinant DNA was
used.

\bibliography{free}
\end{document}